\documentclass[prd,aps,floatfix,nofootinbib,11 pt]{revtex4}
\usepackage{amsmath,graphicx,color,epsfig}

\input{tcilatex}

\begin{document}

\title{ The Effect of Quantum Memory on Quantum Games}
\author{M. Ramzan\thanks{%
mramzan@phys.qau.edu.pk}, Ahmad Nawaz, A. H. Toor and M. K. Khan}
\address{Department of Physics Quaid-i-Azam University \\
Islamabad 45320, Pakistan}

\begin{abstract}
We study quantum games with correlated noise through a generalized
quantization scheme. We investigate the effects of memory on quantum games,
such as Prisoner's Dilemma, Battle of the Sexes and Chicken, through three
prototype quantum-correlated channels. It is shown that the quantum player
enjoys an advantage over the classical player for all nine cases considered
in this paper for the maximally entangled case. However, the quantum player
can also outperform the classical player for subsequent cases that can be
noted in the case of the Battle of the Sexes game. It can be seen that the
Nash equilibria do not change for all the three games under the effect of
memory.\newline
\end{abstract}

\pacs{102.50.Le, 03.65.Ud, 05.40.Ca}
\maketitle

Keywords: quantum games; correlated noise; quantum channels with memory.%
\newline

\vspace*{1.0cm}

\vspace*{1.0cm}



\section{Introduction}

The study of quantum games combines the laws of quantum mechanics with game
theory. It is interesting to study the games at microscopic level where the
laws of quantum mechanics dictates the dynamics. Quantum games offer
additional strategies to the players and resolve dilemmas that occur in
classical games [1--6]. Quantum theory has already been applied to a wide
variety of games [7--11] and shown to be experimentally feasible [12].
Additionally, quantum games offer a new paradigm for exploring the
fascinating world of quantum information [13--15]. Meyer [16] has also
pointed out the connection between quantum games and quantum information
processing. In the earlier work on quantum games, for simplicity, the role
of channels is mostly ignored. In a realistic setup, however, the flow of
information between players and arbiter is subject to interaction with the
environment. Quantum entanglement, which is one of the interesting features
of quantum mechanics, plays a crucial role in quantum information
processing. When quantum information processing is performed in the real
world, decoherence caused by an external environment is inevitable. In other
words, the influence of an external environmental system on the entanglement
cannot be ignored. Recently, decoherence effects in quantum games have been
studied [17].

Later, interest has been developed to extend the theory of quantum channels
to encompass memory effects [18, 19]. There are time scales for which
successive uses of channel are correlated and memory effects need to be
taken into account. Quantum computing in the presence of noise is possible
with the use of decoherence-free subspaces [20] and quantum error correction
[21]. Studies concerning quantum games in the presence of decoherence and
correlated noise have produced interesting results. Chen et al [17] have
shown that in the case of the game Prisoner's Dilemma, the Nash equilibria
are not changed by the effects of decoherence for maximally entangled states
incorporating three prototype decoherence channels. Recently, Nawaz and Toor
[22] have shown for the quantum games based on quantum-correlated
phase-damping channel that the quantum player only enjoys an advantage over
the classical player when both the initial quantum state and the measurement
basis are in entangled form. It is also shown that for maximum correlation
the effects of decoherence diminish and it behaves as a noiseless game.
Recently, Cao et al [23] have investigated the effect of quantum noise on a
multiplayer quantum game. They have shown that in a maximally entangled case
a special Nash equilibrium appears for a specific range of quantum noise
parameter.

In this paper, we study the quantum games based on three prototype quantum
correlated channels (QCC) parameterized by a memory factor $%
\mu
$ which measures the degree of correlations, in the context of generalized
quantization scheme for non-zero sum games [24]. We identify four different
regimes on the basis of initial state and measurement basis entanglement
parameters, $\gamma \in \lbrack 0,\pi /2]$ and $\delta \in \lbrack 0,\pi /2]$%
, respectively. For these four regimes, we study the role of decoherence
parameter $p\in \lbrack 0,1]$ and memory parameter $\mu \in \lbrack 0,1]$
for three quantum games. Here, $\delta =0$ means that the measurement basis
are unentangled and $\delta =\pi /2$ means that it is maximally entangled, $%
\gamma =0$ means that the game is initially unentangled and $\gamma =\pi /2$
means that it is maximally entangled. Whereas the lower and upper limits of $%
p$ correspond to a fully coherent and fully decohered system, respectively.
Furthermore, the lower and upper limits of $\mu $ correspond to a memoryless
and maximum memory (degree of correlation) cases, respectively. It is shown
that for $\gamma =\delta =0$, with decoherence and noise parameters $%
p_{1}=p_{2}=0$ and $\mu _{1}=\mu _{2}=0,$ respectively, the game reduces to
the classical one for all the cases discussed in this paper. In Prisoner's
Dilemma game, when $\gamma \neq $ $0$, $\delta $ $=0$, it is interesting to
note that though the initial state is entangled, the quantum player has no
advantage over the classical player in Prisoner's Dilemma and Chicken games.
The same happens for the case of $\gamma =0,$ $\delta \neq 0.$ An
interesting aspect of these cases arises based on entangling parameter $%
\gamma $ and measurement parameter $\delta $ for $\delta $ $=0$, $\gamma
\neq $ $0$ and $\gamma =0,$ $\delta \neq 0$ in Battle of Sexes game. It is
seen that the quantum player is better off for both of the above cases for $%
p>0$ in case of amplitude-damping and depolarizing channels respectively.
For the case when $\gamma =$ $\delta $ $=$ $\pi /2$ , the quantum player
always remains better off for all values of $p$ against a player restricted
to classical strategies for all the nine cases considered.

\section{Quantum channels with memory}

Several investigations concern the transmission of quantum information from
one party (Alice) to another (Bob) through a communication channel. In the
most basic configuration the information is encoded in qubits. If the qubits
are perfectly protected from environmental influence, Bob receives them in
the same state prepared by Alice. In the more realistic case, however, the
qubits have a nontrivial dynamics during the transmission because of their
interaction with the environment [25]. Therefore, Bob receives a set of
distorted qubits because of the disturbing action of the channel. Recently,
the study of quantum channels has attracted a lot of attention [18, 19, 26].
Early works in this direction were devoted, mainly, to memoryless channels
for which consecutive signal transmissions through the channel are not
correlated. Correlated noise, also referred as memory in the literature,
acts on consecutive uses of the channels. However in general one may want to
encode classical data into entangled strings or, consecutive uses of the
channel may be correlated to each other. Hence, we are dealing with a
strongly correlated quantum system, the correlation of which results from
the memory of the channel itself. In our model Alice and Bob, each uses
individual channels to communicate with the arbiter of the game. Alice's
channel is correlated in time (and therefore has a memory), i.e. the two
uses of the channel; the first passage (from the arbiter) and the second
passage (back to the arbiter) through the channel are correlated. A similar
situation occurs for Bob as depicted in figure 1. We consider here different
noise models based on phase-damping, amplitude-damping and depolarizing
channels.

The action of transmission channels is described by Kraus operators which
satisfy $\sum\limits_{i=0}^{1}A_{i}^{\dagger }A_{i}=1.$ In operator sum
representation the dephasing process can be expressed as [25]. 
\begin{equation}
\rho _{f}=\sum\limits_{i=0}^{1}A_{i}\rho _{in}A_{i}^{\dagger }
\end{equation}%
where $\rho _{in}$ represents the initial density matrix for quantum state
and 
\begin{eqnarray}
A_{0} &=&\sqrt{1-\frac{p}{2}}I  \notag \\
A_{1} &=&\sqrt{\frac{p}{2}}\sigma _{z}
\end{eqnarray}%
are the Kraus operators, $I$ is the identity operator, $p$ is the
decoherence parameter and $\sigma _{z}$ is the Pauli matrix. Let $N$ qubits
are allowed to pass through such a channel then equation (1) becomes [27] 
\begin{equation}
\rho _{f}=\sum\limits_{k_{1,}....,.k_{n}=0}^{1}(A_{k_{n}}\otimes
.....A_{k_{1}})\rho _{in}(A_{k_{1}}^{\dagger }\otimes
.....A_{k_{n}}^{\dagger })
\end{equation}%
Now if the noise is correlated with memory of degree $\mu ,$ then the action
of the channel on two consecutive qubits is given by Kraus operator [18]%
\begin{equation}
A_{ij}=\sqrt{p_{i}[(1-\mu )p_{j}+\mu \delta _{ij}]}\sigma _{i}\otimes \sigma
_{j}
\end{equation}%
where $\sigma _{i}$ and $\sigma _{j}$ are usual Pauli matrices with indices $%
i$ and $j$ running from $0$ to $3.$ The above expression means that with the
probability $1-%
\mu
$ the noise is uncorrelated whereas with probability $%
\mu
$ the noise is correlated as illustrated in the below equations. Physically
the parameter $%
\mu
$ is determined by the relaxation time of the channel when a qubit passes
through it. In order to remove correlations, one can wait until the channel
has relaxed to its original state before sending the next qubit, however
this lowers the rate of information transfer. Thus it is necessary to
consider the performance of the channel for arbitrary values of $%
\mu
$ to reach a compromise between various factors which determine the final
rate of information transfer.\ Thus in passing through the channel any two
consecutive qubits undergo random independent (uncorrelated) errors with
probability $1-%
\mu
$ and identical (correlated) errors with probability $%
\mu
$. This should be the case if the channel has a memory depending on its
relaxation time and if we stream the qubits through it.\ A quantum dephasing
channel (Pauli $Z$ channel) with uncorrelated noise (memoryless channel) can
be defined as one specified by the following Kraus operators 
\begin{equation}
Z_{ij}^{u}=\sqrt{p_{i}p_{j}}\sigma _{i}\otimes \sigma _{j},\qquad i,j=0,3
\end{equation}%
and one with correlated noise (channel with memory) by 
\begin{equation}
Z_{kk}^{c}=\sqrt{p_{k}}\sigma _{k}\otimes \sigma _{k},\qquad k=0,3
\end{equation}%
The action of a depolarizing channel with memory can be expressed as%
\begin{equation}
\pi \rightarrow \rho =\Phi (\pi )=(1-\mu
)\sum\limits_{i,j=0}^{3}D_{ij}^{u}\pi D_{ij}^{u\dagger }+\mu
\sum\limits_{k=0}^{1}D_{kk}^{c}\pi D_{kk}^{c\dagger }
\end{equation}%
where $0\leq \mu \leq 1.$With probability $1-%
\mu
$ the noise is uncorrelated and completely specified by the Kraus operators 
\begin{equation}
D_{ij}^{u}=\sqrt{p_{i}p_{j}}\sigma _{i}\otimes \sigma _{j},
\end{equation}%
and one with correlated noise (channel with memory) by 
\begin{equation}
D_{kk}^{c}=\sqrt{p_{k}}\sigma _{k}\otimes \sigma _{k},
\end{equation}%
where $0\leq p\leq 1,$ $p_{0}=(1-p),$ $p_{1}=p_{2}=p_{3}=p/3.$ However, we
note that a quantum amplitude-damping channel with uncorrelated noise can be
defined as one specified by the following Kraus operators: 
\begin{equation}
A_{00}^{u}=A_{0}\otimes A_{0},\qquad A_{01}^{u}=A_{0}\otimes A_{1},\qquad
A_{10}^{u}=A_{1}\otimes A_{0},\qquad A_{11}^{u}=A_{1}\otimes A_{1}
\end{equation}%
\begin{equation}
A_{0}=\left[ 
\begin{array}{cc}
\cos \chi & 0 \\ 
0 & 1%
\end{array}%
\right] ,\ \ \ A_{1}=\left[ 
\begin{array}{cc}
0 & 0 \\ 
\sin \chi & 0%
\end{array}%
\right]
\end{equation}%
However, the Kraus operators for a quantum amplitude-damping channel with
correlated noise are given by Yeo and Skeen [19] as under:

\begin{equation}
A_{00}^{c}=\left[ 
\begin{array}{llll}
\cos \chi & 0 & 0 & 0 \\ 
0 & 1 & 0 & 0 \\ 
0 & 0 & 1 & 0 \\ 
0 & 0 & 0 & 1%
\end{array}%
\right] ,\ \ \ A_{11}^{c}=\left[ 
\begin{array}{llll}
0 & 0 & 0 & 0 \\ 
0 & 0 & 0 & 0 \\ 
0 & 0 & 0 & 0 \\ 
\sin \chi & 0 & 0 & 0%
\end{array}%
\right]
\end{equation}%
where, $0\leq \chi \leq \pi /2$ and is related to decoherence parameter as 
\begin{eqnarray}
\cos ^{2}\chi &=&1-p  \notag \\
\sin ^{2}\chi &=&p
\end{eqnarray}%
It is clear that $A_{00}^{c}$ cannot be written as a tensor product of two
two-by-two matrices. This gives rise to the typical spooky action of the
channel: $\left\vert 01\right\rangle $ and $\left\vert 10\right\rangle $,
and any linear combination of them, and $\left\vert 11\right\rangle $ will
go through the channel undisturbed, but not $\left\vert 00\right\rangle .$%
The action of this non-unital channel is given by 
\begin{equation}
\pi \rightarrow \rho =\Phi (\pi )=(1-\mu
)\sum\limits_{i,j=0}^{1}A_{ij}^{u}\pi A_{ij}^{u\dagger }+\mu
\sum\limits_{k=0}^{1}A_{kk}^{c}\pi A_{kk}^{c\dagger }
\end{equation}

The protocol for quantum games in the presence of correlated noise is
developed by Nawaz and Toor [22]. We consider that an initial entangled
state is prepared by the arbiter and passed on to the players through three
prototype quantum correlated channels (as shown in figure 1). i.e. Alice and
Bob each uses individual channels to communicate with the arbiter of the
game. Alice's channel is correlated in time (and therefore has a memory),
i.e. the two uses of the channel are correlated. On receiving the quantum
state from the arbiter, the players apply their local operators (strategies)
and return it back to arbiter through QCC. Arbiter then performs the
measurement and announces their payoffs. Let the game start with the initial
quantum state given below,%
\begin{equation}
\left\vert \psi _{in}\right\rangle =\cos \frac{\gamma }{2}\left\vert
00\right\rangle +i\sin \frac{\gamma }{2}\left\vert 11\right\rangle
\end{equation}%
where $0\leq \gamma \leq \pi /2$ corresponds to entanglement of the initial
state. The strategies of the players in the generalized quantization scheme
are represented by the unitary operator $U_{i}$ of the form [24]. 
\begin{equation}
U_{i}=\cos \frac{\theta _{i}}{2}R_{i}+\sin \frac{\theta _{i}}{2}P_{i}
\end{equation}%
where $i=1$ or $2$ and $R_{i}$, $P_{i}$ are the unitary operators defined as 
\begin{eqnarray}
R_{i}\left\vert 0\right\rangle &=&e^{i\alpha _{i}}\left\vert 0\right\rangle
,\qquad \qquad R_{i}\left\vert 1\right\rangle =e^{-i\alpha _{i}}\left\vert
1\right\rangle  \notag \\
P_{i}\left\vert 0\right\rangle &=&e^{i(\frac{\pi }{2}-\beta _{i})}\left\vert
1\right\rangle ,\qquad P_{i}\left\vert 1\right\rangle =e^{i(\frac{\pi }{2}%
+\beta _{i})}\left\vert 0\right\rangle
\end{eqnarray}%
where $0\leq \theta \leq \pi $ and $-\pi \leq \alpha ,$ $\beta \leq \pi .$
Under the generalized quantization scheme with three parameter strategies,
the initial state given in equation (15) transforms to 
\begin{equation}
\rho _{f}=(U_{1}\otimes U_{2})\rho _{in}(U_{1}\otimes U_{2})^{\dagger }
\end{equation}%
where $\rho _{in}=\left\vert \psi _{in}\right\rangle \left\langle \psi
_{in}\right\vert $ is the density matrix for the quantum state. The
operators used by the arbiter to determine the payoff for Alice and Bob are 
\begin{equation}
P=\$_{00}P_{00}+\$_{01}P_{01}+\$_{10}P_{10}+\$_{11}P_{11}
\end{equation}%
where 
\begin{eqnarray}
P_{00} &=&\left\vert \psi _{00}\right\rangle \left\langle \psi
_{00}\right\vert ,\qquad \left\vert \psi _{00}\right\rangle =\cos \frac{%
\delta }{2}\left\vert 00\right\rangle +i\sin \frac{\delta }{2}\left\vert
11\right\rangle  \notag \\
P_{11} &=&\left\vert \psi _{11}\right\rangle \left\langle \psi
_{11}\right\vert ,\qquad \left\vert \psi _{11}\right\rangle =\cos \frac{%
\delta }{2}\left\vert 11\right\rangle +i\sin \frac{\delta }{2}\left\vert
00\right\rangle  \notag \\
P_{10} &=&\left\vert \psi _{10}\right\rangle \left\langle \psi
_{10}\right\vert ,\qquad \left\vert \psi _{10}\right\rangle =\cos \frac{%
\delta }{2}\left\vert 10\right\rangle -i\sin \frac{\delta }{2}\left\vert
01\right\rangle  \notag \\
P_{01} &=&\left\vert \psi _{01}\right\rangle \left\langle \psi
_{01}\right\vert ,\qquad \left\vert \psi _{01}\right\rangle =\cos \frac{%
\delta }{2}\left\vert 01\right\rangle -i\sin \frac{\delta }{2}\left\vert
10\right\rangle
\end{eqnarray}%
with $0\leq \delta \leq \pi /2$ and $\$_{ij}$ are the elements of payoff
matrix in the $i$th row and $j$th column of classical games as given in
appendix A. In the generalized quantization scheme for three set of
parameters, the players payoffs read 
\begin{equation}
\$^{A}(\theta _{i},\alpha _{i},\beta _{i})=\text{Tr}(P_{A}\rho _{f}),\quad
\$^{B}(\theta _{i},\alpha _{i},\beta _{i})=\text{Tr}(P_{B}\rho _{f})
\end{equation}%
where Tr represents the trace of the matrix. Using equations (4)-(9), (14),
(19) and (21), the payoffs of the two players, when both channels (first and
second) are amplitude-damping, are given by%
\begin{eqnarray}
\$(\theta _{i},\alpha _{i},\beta _{i}) &=&c_{1}c_{2}[\eta
_{1}^{A}\$_{00}+\chi _{1}^{A}\$_{11}+\Delta
_{1}^{A}(\$_{01}+\$_{10})+(\$_{11}-\$_{00})\chi _{\mu _{1}}^{(10)}\chi _{\mu
_{2}}^{(10)}\xi \cos 2(\alpha _{1}+\alpha _{2})]  \notag \\
&&+s_{1}s_{2}[\eta _{2}^{A}\$_{00}+\chi _{2}^{A}\$_{11}+\Delta
_{2}^{A}(\$_{01}+\$_{10})-(\$_{11}-\$_{00})\chi _{\mu _{1}}^{(10)}\chi _{\mu
_{2}}^{(10)}\xi \cos 2(\beta _{1}+\beta _{2})]  \notag \\
&&+s_{1}c_{2}[\eta _{3}^{A}\$_{00}+\chi _{3}^{A}\$_{11}+\Delta
_{3}^{A}\$_{01}+\Delta _{4}^{A}\$_{10}+(\$_{01}-\$_{10})\chi _{\mu
_{1}}^{(10)}\chi _{\mu _{2}}^{(b)}\xi \cos 2(\alpha _{2}-\beta _{1})]  \notag
\\
&&+c_{1}s_{2}[\eta _{3}^{A}\$_{00}+\chi _{3}^{A}\$_{11}+\Delta
_{4}^{A}\$_{01}+\Delta _{3}^{A}\$_{10}-(\$_{01}-\$_{10})\chi _{\mu
_{1}}^{(10)}\chi _{\mu _{2}}^{(b)}\xi \cos 2(\alpha _{1}-\beta _{2})]  \notag
\\
&&-\frac{1}{4}[\sin (\delta )\sin (\theta _{1})\sin (\theta _{2})\chi _{\mu
_{2}}^{(10)}(\$_{00}-\$_{11})\Delta _{5}^{A}\sin (\alpha _{1}+\alpha
_{2}+\beta _{1}+\beta _{2})]  \notag \\
&&-\frac{1}{4}[\sin (\gamma )\sin (\theta _{1})\sin (\theta _{2})(\eta
_{4}\$_{00}+\chi _{4}\$_{11})\Delta _{5}^{A}\sin (\alpha _{1}+\alpha
_{2}-\beta _{1}-\beta _{2})]  \notag \\
&&-\frac{1}{4}[\sin (\gamma )\sin (\theta _{1})\sin (\theta
_{2})(\$_{01}+\$_{10})\Delta _{6}^{A}\sin (\alpha _{1}+\alpha _{2}-\beta
_{1}-\beta _{2})]  \notag \\
&&+\frac{1}{4}[\sin (\delta )\sin (\theta _{1})\sin (\theta _{2})\chi _{\mu
_{2}}^{(b)}(\$_{01}-\$_{10})\Delta _{5}^{A}\sin (\alpha _{1}-\alpha
_{2}+\beta _{1}-\beta _{2})]
\end{eqnarray}%
The payoffs of the two players, when both channels are depolarizing, are
given as 
\begin{eqnarray}
\$(\theta _{i},\alpha _{i},\beta _{i}) &=&c_{1}c_{2}[\eta ^{D}\$_{00}+\chi
^{D}\$_{11}+\Delta ^{D}(\$_{01}+\$_{10})+(\$_{00}-\$_{11})(\Delta _{\mu
2}^{4}-\frac{2}{3}\mu _{2}p_{2})\Delta _{\mu 1}^{3}\xi \cos 2(\alpha
_{1}+\alpha _{2})]  \notag \\
&&+s_{1}s_{2}[\chi ^{D}\$_{00}+\eta ^{D}\$_{11}+\Delta
^{D}(\$_{01}+\$_{10})-(\$_{00}-\$_{11})(\Delta _{\mu 2}^{4}+\frac{2}{3}\mu
_{2}p_{2})\Delta _{\mu 1}^{3}\xi \cos 2(\beta _{1}+\beta _{2})]  \notag \\
&&+s_{1}c_{2}[\eta ^{D}\$_{10}+\chi ^{D}\$_{01}+\Delta
^{D}(\$_{00}+\$_{11})+(\$_{01}-\$_{10})(\Delta _{\mu 2}^{4}-\frac{2}{3}\mu
_{2}p_{2})\Delta _{\mu 1}^{3}\xi \cos 2(\alpha _{2}-\beta _{1})]  \notag \\
&&+c_{1}s_{2}[\eta ^{D}\$_{01}+\chi ^{D}\$_{10}+\Delta
^{D}(\$_{00}+\$_{11})-(\$_{01}-\$_{10})(\Delta _{\mu 2}^{4}+\frac{2}{3}\mu
_{2}p_{2})\Delta _{\mu 1}^{3}\xi \cos 2(\alpha _{1}-\beta _{2})]  \notag \\
&&-(\frac{1}{4}\Delta _{\mu 1}^{3}\Delta _{\mu 2}^{1}+\frac{1}{2}\Delta
_{\mu 1}^{3}\Delta _{\mu 2}^{2}-\frac{1}{4}\Delta _{\mu 1}^{3}\Delta _{\mu
2}^{3})(\$_{00}+\$_{11})\sin (\gamma )\sin (\theta _{1})\sin (\theta
_{2})\sin (\alpha _{1}+\alpha _{2}-\beta _{1}-\beta _{2})  \notag \\
&&-(\frac{1}{4}\Delta _{\mu 2}^{4}-\frac{1}{6}\mu
_{2}p_{2})(\$_{00}-\$_{11})\eta _{1DP}\sin (\delta )\sin (\theta _{1})\sin
(\theta _{2})\sin (\alpha _{1}+\alpha _{2}+\beta _{1}+\beta _{2})  \notag \\
&&+(\frac{1}{4}\Delta _{\mu 1}^{3}\Delta _{\mu 2}^{1}-\frac{1}{2}\Delta
_{\mu 1}^{3}\Delta _{\mu 2}^{2}+\frac{1}{4}\Delta _{\mu 1}^{3}\Delta _{\mu
2}^{3})(\$_{01}+\$_{10})\sin (\gamma )\sin (\theta _{1})\sin (\theta
_{2})\sin (\alpha _{1}+\alpha _{2}-\beta _{1}-\beta _{2})  \notag \\
&&-(\frac{1}{4}\Delta _{\mu 2}^{4}-\frac{1}{6}\mu
_{2}p_{2})(\$_{01}-\$_{10})\eta _{1DP}\sin (\delta )\sin (\theta _{1})\sin
(\theta _{2})\sin (\alpha _{1}-\alpha _{2}+\beta _{1}-\beta _{2})]
\end{eqnarray}

The payoffs of the two players, when both channels are phase-damping, are
given by 
\begin{eqnarray}
\$(\theta _{i},\alpha _{i},\beta _{i}) &=&c_{1}c_{2}[\eta ^{P}\$_{00}+\chi
^{P}\$_{11}+(\$_{00}-\$_{11})\mu _{p}^{(1)}\mu _{p}^{(2)}\xi \cos 2(\alpha
_{1}+\alpha _{2})]  \notag \\
&&+s_{1}s_{2}[\eta ^{P}\$_{11}+\chi ^{P}\$_{00}-(\$_{00}-\$_{11})\mu
_{p}^{(1)}\mu _{p}^{(2)}\xi \cos 2(\beta _{1}+\beta _{2})]  \notag \\
&&+s_{1}c_{2}[\eta ^{P}\$_{10}+\chi ^{P}\$_{01}+(\$_{10}-\$_{01})\mu
_{p}^{(1)}\mu _{p}^{(2)}\xi \cos 2(\alpha _{2}-\beta _{1})]  \notag \\
&&+c_{1}s_{2}[\eta ^{P}\$_{01}+\chi ^{P}\$_{10}-(\$_{10}-\$_{01})\mu
_{p}^{(1)}\mu _{p}^{(2)}\xi \cos 2(\alpha _{1}-\beta _{2})]  \notag \\
&&+\frac{\mu _{p}^{(2)}}{4}(\$_{00}-\$_{11})\sin (\delta )\sin (\theta
_{1})\sin (\theta _{2})\sin (\alpha _{1}+\alpha _{2}+\beta _{1}+\beta _{2}) 
\notag \\
&&+\frac{\mu _{p}^{(1)}}{4}(-\$_{00}-\$_{11}+\$_{01}+\$_{10})\sin (\gamma
)\sin (\theta _{1})\sin (\theta _{2})\sin (\alpha _{1}+\alpha _{2}-\beta
_{1}-\beta _{2})  \notag \\
&&+\frac{\mu _{p}^{(2)}}{4}(\$_{10}-\$_{01})\sin (\delta )\sin (\theta
_{1})\sin (\theta _{2})\sin (\alpha _{1}-\alpha _{2}+\beta _{1}-\beta _{2})
\end{eqnarray}%
The payoffs of the two players, when first channel is phase-damping and
second channel is amplitude-damping, are given by 
\begin{eqnarray}
\$(\theta _{i},\alpha _{i},\beta _{i}) &=&c_{1}c_{2}[\eta
_{1}^{PA}\$_{00}+\chi _{1}^{PA}\$_{11}+\Delta
_{1}^{PA}(\$_{01}+\$_{10})+(\$_{00}-\$_{11})\mu _{p}^{(1)}\chi _{\mu
2}^{(10)}\xi \cos 2(\alpha _{1}+\alpha _{2})]  \notag \\
&&+s_{1}s_{2}[\eta _{2}^{PA}\$_{00}+\chi _{2}^{PA}\$_{11}+\Delta
_{2}^{PA}(\$_{01}+\$_{10})-(\$_{00}-\$_{11})\mu _{p}^{(1)}\chi _{\mu
2}^{(10)}\xi \cos 2(\beta _{1}+\beta _{2})]  \notag \\
&&+s_{1}c_{2}[\Delta _{3}^{PA}\$_{01}+\Delta _{4}^{PA}\$_{10}+\eta
_{3}^{PA}\$_{00}+\chi _{3}^{PA}\$_{11}+(\$_{01}-\$_{10})\mu _{p}^{(1)}\chi
_{\mu 2}^{(b)}\xi \cos 2(\alpha _{2}-\beta _{1})]  \notag \\
&&+c_{1}s_{2}[\Delta _{4}^{PA}\$_{01}+\Delta _{3}^{PA}\$_{10}+\eta
_{3}^{PA}\$_{00}+\chi _{3}^{PA}\$_{11}-(\$_{01}-\$_{10})\mu _{p}^{(1)}\chi
_{\mu 2}^{(b)}\xi \cos 2(\alpha _{1}-\beta _{2})]  \notag \\
&&-\frac{1}{4}(\chi _{\mu 2}^{(00)}+\chi _{\mu 2}^{(11)}-2\chi _{\mu
2}^{(a)})(\$_{00}-\$_{11})\mu _{p}^{(1)}\sin (\gamma )\sin (\theta _{1})\sin
(\theta _{2})\sin (\alpha _{1}+\alpha _{2}-\beta _{1}-\beta _{2})  \notag \\
&&+\frac{1}{4}\chi _{\mu 2}^{(10)}(\$_{00}-\$_{11})\sin (\delta )\sin
(\theta _{1})\sin (\theta _{2})\sin (\alpha _{1}+\alpha _{2}+\beta
_{1}+\beta _{2})  \notag \\
&&+\frac{1}{4}\mu _{p}^{(1)}(\chi _{\mu 2}^{(b)}+\chi _{\mu
2}^{(01)})(\$_{01}+\$_{10})\sin (\gamma )\sin (\theta _{1})\sin (\theta
_{2})\sin (\alpha _{1}+\alpha _{2}-\beta _{1}-\beta _{2})  \notag \\
&&+\frac{1}{4}\chi _{\mu 2}^{(b)}(\$_{01}-\$_{10})\sin (\delta )\sin (\theta
_{1})\sin (\theta _{2})\sin (\alpha _{1}-\alpha _{2}+\beta _{1}-\beta _{2})
\end{eqnarray}%
The payoffs of the two players, when first channel is amplitude-damping and
second channel is phase-damping, are given by 
\begin{eqnarray}
\$(\theta _{i},\alpha _{i},\beta _{i}) &=&c_{1}c_{2}[\eta ^{AP}\$_{00}+\chi
^{AP}\$_{11}+\Delta ^{AP}(\$_{01}+\$_{10})+(\$_{00}-\$_{11})\mu
_{p}^{(2)}\chi _{\mu 1}^{(10)}\xi \cos 2(\alpha _{1}+\alpha _{2})]  \notag \\
&&+s_{1}s_{2}[\chi ^{AP}\$_{00}+\eta ^{AP}\$_{11}+\Delta
^{AP}(\$_{01}+\$_{10})-(\$_{00}-\$_{11})\mu _{p}^{(2)}\chi _{\mu
1}^{(10)}\xi \cos 2(\beta _{1}+\beta _{2})]  \notag \\
&&+s_{1}c_{2}[\eta ^{AP}\$_{10}+\chi ^{AP}\$_{01}+\Delta
^{AP}(\$_{00}+\$_{11})+(\$_{01}-\$_{10})\mu _{p}^{(2)}\chi _{\mu
1}^{(10)}\xi \cos 2(\alpha _{2}-\beta _{1})]  \notag \\
&&+c_{1}s_{2}[\eta ^{AP}\$_{01}+\chi ^{AP}\$_{10}+\Delta
^{AP}(\$_{00}+\$_{11})-(\$_{01}-\$_{10})\mu _{p}^{(2)}\chi _{\mu
1}^{(10)}\xi \cos 2(\alpha _{1}-\beta _{2})]  \notag \\
&&-\frac{1}{4}\chi _{\mu 1}^{(10)}(\$_{00}+\$_{11})\sin (\gamma )\sin
(\theta _{1})\sin (\theta _{2})\sin (\alpha _{1}+\alpha _{2}-\beta
_{1}-\beta _{2})  \notag \\
&&+\frac{1}{4}\mu _{p}^{(2)}\eta _{1AD}(\$_{00}+\$_{11})\sin (\delta )\sin
(\theta _{1})\sin (\theta _{2})\sin (\alpha _{1}+\alpha _{2}+\beta
_{1}+\beta _{2})  \notag \\
&&+\frac{1}{4}\chi _{\mu 1}^{(10)}(\$_{01}+\$_{10})\sin (\gamma )\sin
(\theta _{1})\sin (\theta _{2})\sin (\alpha _{1}+\alpha _{2}-\beta
_{1}-\beta _{2})  \notag \\
&&+\frac{1}{4}\mu _{p}^{(2)}\eta _{1AD}(\$_{01}-\$_{10})\sin (\delta )\sin
(\theta _{1})\sin (\theta _{2})\sin (\alpha _{1}-\alpha _{2}+\beta
_{1}-\beta _{2})
\end{eqnarray}%
The payoffs of the two players, when first channel is amplitude-damping and
second channel is depolarizing, are given by 
\begin{eqnarray}
\$(\theta _{i},\alpha _{i},\beta _{i}) &=&c_{1}c_{2}[\eta ^{AD}\$_{00}+\chi
^{AD}\$_{11}+\Delta ^{AD}(\$_{01}+\$_{10})+(\$_{00}-\$_{11})(\Delta _{\mu
2}^{4}-\frac{2}{3}\mu _{2}p_{2})\chi _{\mu 1}^{(10)}\xi \cos 2(\alpha
_{1}+\alpha _{2})]  \notag \\
&&+s_{1}s_{2}[\chi ^{AD}\$_{00}+\eta ^{AD}\$_{11}+\Delta
^{AD}(\$_{01}+\$_{10})-(\$_{00}-\$_{11})(\Delta _{\mu 2}^{4}-\frac{2}{3}\mu
_{2}p_{2})\chi _{\mu 1}^{(10)}\xi \cos 2(\alpha _{1}+\alpha _{2})]  \notag \\
&&+s_{1}c_{2}[\chi ^{AD}\$_{01}+\eta ^{AD}\$_{10}+\Delta
^{AD}(\$_{00}+\$_{11})+(\$_{01}-\$_{10})(\Delta _{\mu 2}^{4}-\frac{2}{3}\mu
_{2}p_{2})\chi _{\mu 1}^{(10)}\xi \cos 2(\alpha _{2}-\beta _{1})]  \notag \\
&&+c_{1}s_{2}[\eta ^{AD}\$_{01}+\chi ^{AD}\$_{10}+\Delta
^{AD}(\$_{00}+\$_{11})-(\$_{01}-\$_{10})(\Delta _{\mu 2}^{4}-\frac{2}{3}\mu
_{2}p_{2})\chi _{\mu 1}^{(10)}\xi \cos 2(\alpha _{1}-\beta _{2})]  \notag \\
&&-(\frac{1}{4}\chi _{\mu 1}^{(10)}\Delta _{\mu 2}^{1}-\frac{1}{2}\chi _{\mu
1}^{(10)}\Delta _{\mu 2}^{2}+\frac{1}{4}\chi _{\mu 1}^{(10)}\Delta _{\mu
2}^{3})(\$_{00}+\$_{11})\sin (\gamma )\sin (\theta _{1})\sin (\theta
_{2})\sin (\alpha _{1}+\alpha _{2}-\beta _{1}-\beta _{2})  \notag \\
&&+(\frac{1}{4}\Delta _{\mu 2}^{4}-\frac{1}{6}\mu
_{2}p_{2})(\$_{00}-\$_{11})\eta _{1AD}\sin (\delta )\sin (\theta _{1})\sin
(\theta _{2})\sin (\alpha _{1}+\alpha _{2}+\beta _{1}+\beta _{2})  \notag \\
&&+(\frac{1}{4}\chi _{\mu 1}^{(10)}\Delta _{\mu 2}^{1}-\frac{1}{2}\chi _{\mu
1}^{(10)}\Delta _{\mu 2}^{2}+\frac{1}{4}\chi _{\mu 1}^{(10)}\Delta _{\mu
2}^{3})(\$_{01}+\$_{10})\sin (\gamma )\sin (\theta _{1})\sin (\theta
_{2})\sin (\alpha _{1}+\alpha _{2}-\beta _{1}-\beta _{2})  \notag \\
&&+(\frac{1}{4}\Delta _{\mu 2}^{4}-\frac{1}{6}\mu
_{2}p_{2})(\$_{01}-\$_{10})\eta _{1AD}\sin (\delta )\sin (\theta _{1})\sin
(\theta _{2})\sin (\alpha _{1}-\alpha _{2}+\beta _{1}-\beta _{2})]
\end{eqnarray}%
The payoffs of the two players, when first channel is depolarizing and
second channel is amplitude-damping, are given by 
\begin{eqnarray}
\$(\theta _{i},\alpha _{i},\beta _{i}) &=&c_{1}c_{2}[\eta
_{1}^{DA}\$_{00}+\chi _{1}^{DA}\$_{11}+\Delta
_{1}^{DA}(\$_{01}+\$_{10})+(\$_{00}-\$_{11})\Delta _{\mu 1}^{3}\chi _{\mu
2}^{(10)}\chi _{\mu 1}^{(10)}\xi \cos 2(\alpha _{1}+\alpha _{2})]  \notag \\
&&+s_{1}s_{2}[\eta _{2}^{DA}\$_{00}+\chi _{2}^{DA}\$_{11}+\Delta
_{1}^{DA}(\$_{01}+\$_{10})-(\$_{00}-\$_{11})\Delta _{\mu 1}^{3}\chi _{\mu
2}^{(10)}\chi _{\mu 1}^{(10)}\xi \cos 2(\beta _{1}+\beta _{2})]  \notag \\
&&+s_{1}c_{2}[\Delta _{2}^{DA}\$_{01}+\Delta _{3}^{DA}\$_{10}+\eta
_{3}^{DA}\$_{00}+\chi _{3}^{DA}\$_{11}+(\$_{01}-\$_{10})\Delta _{\mu
1}^{3}\chi _{\mu 2}^{(b)}\xi \cos 2(\alpha _{2}-\beta _{1})  \notag \\
&&+c_{1}s_{2}[\Delta _{3}^{DA}\$_{01}+\Delta _{2}^{DA}\$_{10}+\eta
_{3}^{DA}\$_{00}+\chi _{3}^{DA}\$_{11}-(\$_{01}-\$_{10})\Delta _{\mu
1}^{3}\chi _{\mu 2}^{(b)}\xi \cos 2(\alpha _{1}-\beta _{2})]  \notag \\
&&-(\frac{1}{4}\chi _{\mu 2}^{(00)}\Delta _{\mu 1}^{3}+\frac{1}{4}\Delta
_{\mu 1}^{3}+\frac{1}{4}\Delta _{\mu 1}^{3}\chi _{\mu
2}^{(11)})(\$_{00}+\$_{11})\sin (\gamma )\sin (\theta _{1})\sin (\theta
_{2})\sin (\alpha _{1}+\alpha _{2}-\beta _{1}-\beta _{2})  \notag \\
&&-(\frac{1}{4}\eta _{1DP}\chi _{\mu 2}^{(10)})(\$_{00}-\$_{11})\sin (\delta
)\sin (\theta _{1})\sin (\theta _{2})\sin (\alpha _{1}+\alpha _{2}+\beta
_{1}+\beta _{2})  \notag \\
&&-(\frac{1}{2}\chi _{\mu 2}^{(01)}\Delta _{\mu 1}^{3}-\frac{1}{4}\chi _{\mu
2}^{(b)}\Delta _{\mu 1}^{3})(\$_{01}+\$_{10})\sin (\gamma )\sin (\theta
_{1})\sin (\theta _{2})\sin (\alpha _{1}+\alpha _{2}-\beta _{1}-\beta _{2}) 
\notag \\
&&-(\frac{1}{4}\eta _{1DP}\chi _{\mu 2}^{(b)})(\$_{01}-\$_{10})\sin (\delta
)\sin (\theta _{1})\sin (\theta _{2})\sin (\alpha _{1}-\alpha _{2}+\beta
_{1}-\beta _{2})]
\end{eqnarray}%
The payoffs of the two players, when first channel is depolarizing and
second channel is phase-damping, are given by 
\begin{eqnarray}
\$(\theta _{i},\alpha _{i},\beta _{i}) &=&c_{1}c_{2}[\eta ^{DP}\$_{00}+\chi
^{DP}\$_{11}+\Delta _{\mu 1}^{4}(\$_{01}+\$_{10})+(\$_{00}-\$_{11})\Delta
_{\mu 1}^{3}\mu _{p}^{2}\xi \cos 2(\alpha _{1}+\alpha _{2})]  \notag \\
&&+s_{1}s_{2}[\chi ^{DP}\$_{00}+\eta ^{DP}\$_{11}+\Delta _{\mu
1}^{4}(\$_{01}+\$_{10})-(\$_{00}-\$_{11})\Delta _{\mu 1}^{3}\mu _{p}^{2}\xi
\cos 2(\beta _{1}+\beta _{2})]  \notag \\
&&+s_{1}c_{2}[\chi ^{DP}\$_{01}+\eta ^{DP}\$_{10}+\Delta _{\mu
1}^{4}(\$_{00}+\$_{11})+(\$_{01}-\$_{10})\Delta _{\mu 1}^{3}\mu _{p}^{2}\xi
\cos 2(\alpha _{2}-\beta _{1})]  \notag \\
&&+c_{1}s_{2}[\chi ^{DP}\$_{10}+\eta ^{DP}\$_{01}+\Delta _{\mu
1}^{4}(\$_{00}+\$_{11})-(\$_{01}-\$_{10})\Delta _{\mu 1}^{3}\mu _{p}^{2}\xi
\cos 2(\alpha _{1}-\beta _{2})]  \notag \\
&&-(\frac{1}{4}\Delta _{\mu 1}^{3})(\$_{00}+\$_{11})\sin (\gamma )\sin
(\theta _{1})\sin (\theta _{2})\sin (\alpha _{1}+\alpha _{2}-\beta
_{1}-\beta _{2})  \notag \\
&&-(\frac{1}{4}\eta _{1DP}\mu _{p}^{2})(\$_{00}-\$_{11})\sin (\delta )\sin
(\theta _{1})\sin (\theta _{2})\sin (\alpha _{1}+\alpha _{2}+\beta
_{1}+\beta _{2})  \notag \\
&&+(\frac{1}{4}\Delta _{\mu 1}^{3})(\$_{01}+\$_{10})\sin (\gamma )\sin
(\theta _{1})\sin (\theta _{2})\sin (\alpha _{1}+\alpha _{2}-\beta
_{1}-\beta _{2})  \notag \\
&&+(\frac{1}{4}\eta _{1DP}\mu _{p}^{2})(\$_{01}-\$_{10})\sin (\delta )\sin
(\theta _{1})\sin (\theta _{2})\sin (\alpha _{1}-\alpha _{2}+\beta
_{1}-\beta _{2})]
\end{eqnarray}%
The payoffs of the two players, when first channel is phase-damping and
second channel is depolarizing, are given by

\begin{eqnarray}
\$(\theta _{i},\alpha _{i},\beta _{i}) &=&c_{1}c_{2}[\eta ^{PD}\$_{00}+\chi
^{PD}\$_{11}+\Delta _{\mu 2}^{2}(\$_{01}+\$_{10})+(\$_{00}-\$_{11})(\Delta
_{\mu 2}^{4}-\frac{2}{3}\mu _{2}p_{2})\mu _{p}^{1}\xi \cos 2(\alpha
_{1}+\alpha _{2})]  \notag \\
&&+s_{1}s_{2}[\chi ^{PD}\$_{00}+\eta ^{PD}\$_{11}+\Delta _{\mu
2}^{2}(\$_{01}+\$_{10})-(\$_{00}-\$_{11})(\Delta _{\mu 2}^{4}-\frac{2}{3}\mu
_{2}p_{2})\mu _{p}^{1}\xi \cos 2(\beta _{1}+\beta _{2})]  \notag \\
&&+s_{1}c_{2}[\chi ^{PD}\$_{01}+\eta ^{PD}\$_{10}+\Delta _{\mu
2}^{1}(\$_{00}+\$_{11})+(\$_{01}-\$_{10})(\Delta _{\mu 2}^{4}-\frac{2}{3}\mu
_{2}p_{2})\mu _{p}^{1}\xi \cos 2(\alpha _{2}-\beta _{1})]  \notag \\
&&+c_{1}s_{2}[\chi ^{PD}\$_{10}+\eta ^{PD}\$_{01}+\Delta _{\mu
2}^{1}(\$_{00}+\$_{11})-(\$_{01}-\$_{10})(\Delta _{\mu 2}^{4}-\frac{2}{3}\mu
_{2}p_{2})\mu _{p}^{1}\xi \cos 2(\alpha _{1}-\beta _{2})]  \notag \\
&&-(\frac{1}{4}\mu _{p}^{1}\Delta _{\mu 2}^{1}-\frac{1}{2}\mu _{p}^{1}\Delta
_{\mu 2}^{2}+\frac{1}{4}\mu _{p}^{1}\Delta _{\mu
2}^{3})(\$_{00}+\$_{11})\sin (\gamma )\sin (\theta _{1})\sin (\theta
_{2})\sin (\alpha _{1}+\alpha _{2}-\beta _{1}-\beta _{2})  \notag \\
&&+(\frac{1}{4}\Delta _{\mu 2}^{4}-\frac{1}{6}\mu _{p}^{1}\mu
_{2}p_{2})(\$_{00}-\$_{11})\sin (\delta )\sin (\theta _{1})\sin (\theta
_{2})\sin (\alpha _{1}+\alpha _{2}+\beta _{1}+\beta _{2})  \notag \\
&&+(\frac{1}{4}\mu _{p}^{1}\Delta _{\mu 2}^{1}-\frac{1}{2}\mu _{p}^{1}\Delta
_{\mu 2}^{2}+\frac{1}{4}\mu _{p}^{1}\Delta _{\mu
2}^{3})(\$_{01}+\$_{10})\sin (\gamma )\sin (\theta _{1})\sin (\theta
_{2})\sin (\alpha _{1}+\alpha _{2}-\beta _{1}-\beta _{2})  \notag \\
&&+(\frac{1}{4}\Delta _{\mu 2}^{4}-\frac{1}{6}\mu _{p}^{1}\mu
_{2}p_{2})(\$_{01}-\$_{10})\sin (\delta )\sin (\theta _{1})\sin (\theta
_{2})\sin (\alpha _{1}-\alpha _{2}+\beta _{1}-\beta _{2})]
\end{eqnarray}%
The definitions of the parameters in the payoffs for equations (22) to (30)
are given in appendix B.

The payoff for the two players can be found by substituting the appropriate
values for $\$_{ij}$ (elements of payoff matrix for the corresponding game
as given in appendix A) in the above equations. These payoffs become the
classical payoffs for $\gamma =\delta =0$ and $p_{1}=p_{2}=0.$ It can be
easily proved that for $\gamma =$ $\delta $ $=$ $\pi /2,$ with $\beta _{1}=$ 
$\beta _{2}=0,$ $\mu _{1}=\mu _{2}=0$ and $p_{1}=1$ or $p_{2}=1$, the
results of ref. [17] are reproduced for all the nine cases in Prisoner's
Dilemma game. Nawaz and Toor have shown that in case of phase-damping
channel, for maximum correlation the effects of decoherence diminish and it
behaves as a noiseless game [22]. However, in case of amplitude and
depolarizing channels, for maximum correlation the effects of decoherence
persist and causes a reduction in the payoffs and it does not behave as a
noiseless game.

\section{Results and discussions}

To analyze the effects of memory in quantum games, we consider a situation
in which Alice is restricted to play classical strategies, i.e., $\alpha
_{1}=\beta _{1}=0$, whereas Bob is allowed to play the quantum strategies as
well. Under these circumstances following four cases for the different
combinations of $\delta $ and $\gamma $ are worth noting;\newline
\textbf{Case(i):} When $\delta =$ $\gamma =0,$ the payoffs reduce to
classical results for unital case i.e. phase-damping and depolarizing
channels. These payoffs, as expected, are independent of the quantum
strategies $\alpha _{2},$ $\beta _{2}$, but only depend upon decoherence
parameter $p$ and the memory parameter $\mu $. For non-unital case, i.e.
amplitude-damping channel, the results\ reduce to classical game when we put 
$p_{1}=p_{2}=0$ along with $\mu _{1}=\mu _{2}=0.$ However, the payoffs of
the two players remain independent of quantum phases and the decrease due to
decoherence is compensated by the memory and payoffs are enhanced from their
classical counterparts (which can be seen from figure 2 for all the three
games).\newline
\textbf{Case(ii):} When $\delta =0,$ $\gamma \neq 0,$ and channels 1 \& 2
are amplitude-damping;

\textbf{a)} In case of Prisoner's Dilemma and Chicken games, the effect of
memory can be summarized as; when $p$ increases the payoffs start
decreasing, however, this effect is partially overcome by the addition of
memory i.e. as $\mu $ increases the payoff increases and as result it
compensates the reduction in player's payoffs due to decoherence (as shown
in figure 3).

\textbf{b)} In case of Battle of Sexes game, the quantum player enjoys an
advantage over classical player for $0<p\leq 1$ (it can be seen from figure
4). The optimal strategy for Bob is to play $\alpha _{2}=\pi /2$ and $\beta
_{2}=0.$

\textbf{c)} When channels 1 \& 2 are phase-damping and amplitude-damping or
depolarizing and amplitude-damping respectively, the quantum player remains
superior over the classical player in case of Battle of Sexes game only
(which can be seen from figure 4).

\textbf{d)} When channels 1 \& 2 are phase-damping or depolarizing, the
payoffs of the players remains equal in all the three games, however, memory
controls the payoffs reduction due to decoherence.\newline
\textbf{Case(iii):} When $\gamma =0,$ $\delta $ $\neq 0,$ and channels 1 \&
2 are depolarizing,

\textbf{a)} In case of Battle of Sexes game, the quantum player outperforms
the classical player for $0<p\leq 1$ (it can be seen from figure 5). The
optimal strategy for Bob is to play$\ \alpha _{2}=0$ and $\beta _{2}=\pi /2.$

\textbf{b)} For phase-damping or amplitude-damping channels, the payoffs of
the players remain equal in all the three games considered and memory
compensates the decoherence effects in Prisoner's Dilemma, and Chicken games.

\textbf{c)} It can be seen from figure 5 that when channels 1 \& 2 are
phase-damping and amplitude-damping or depolarizing and amplitude-damping
respectively, the quantum player remains superior over the classical player
in case of Battle of Sexes game.\newline
\textbf{Case(iv):} when $\gamma =\delta $ $=\pi /2,$ with $\mu =0$
(memoryless case)$,$ the quantum player is better off for $p<1$ for all the
three channels$.$ For $\mu \neq 0$, the quantum player outperform classical
player even for maximum noise, i.e., $p=1$, for all the nine cases, which is
not possible in memoryless case (it can be seen from figures 6 and 7) for
amplitude-damping and depolarizing channels, the similar behaviour is seen
for all the remaining 7 channels).

A Nash equilibrium implies that no player can increase his/her payoff by
unilaterally changing his/her strategy. One can see from case (ii)-b that
for Alice $\theta _{1}=0$ and for Bob $\theta _{2}=\pi /2$ and $\alpha
_{2}=\pi /2$, $\beta _{2}=0$ remain their best strategies throughout the
course of the game for the entire range of the decoherence parameter $p$ and
the memory parameter $\mu $. Similarly, for case (iii)-a, it can be seen
that for Alice $\theta _{1}=0$ and for Bob $\theta _{2}=\pi /2$ and $\alpha
_{2}=0$, $\beta _{2}=\pi /2$ remain their best strategies for all values of $%
p$ and $\mu $ and no player can increase his/her payoff by unilaterally
changing his/her strategy. A similar situation occurs for all the remaining
cases. Thus by inspection (from equations (22) to (30)), one can see that
the Nash equilibria of the three games do not change under the effect of
quantum memory.

\section{Conclusions}

Quantum games with correlated noise are studied under the generalized
quantization scheme [24]. Three games, Prisoner's Dilemma, Battle of Sexes
and Chicken are studied with one player restricted to the classical
strategies while the other is allowed to play quantum strategies. It is
shown that the effects of the memory and decoherence become effective for
the case $\gamma =\delta $ $=\pi /2$, for which the quantum player out
performs the classical player in all the three games for maximally entangled
case. It is also shown that the quantum player enjoys an advantage over
classical player for $\delta =0,$ $\gamma \neq 0$ and $\gamma =0,$ $\delta $ 
$\neq 0$ cases in Battle of Sexes game when amplitude-damping \ and
depolarizing channels are used respectively. It can be seen that the Nash
equilibria of the three games do not change under the effect of
memory.\bigskip

\begin{center}
\textbf{Appendix A: Classical Games}
\end{center}

Brief descriptions of three classical games, Prisoner's Dilemma, Battle of
Sexes and Chicken are given below,

\begin{center}
\textbf{Prisoner's Dilemma }
\end{center}

This game depicts a situation where two prisoners, who have committed a
crime together, are being interrogated in separate cells. The two possible
moves for each prisoner are, to cooperate $(C)$ or to defect $(D)$. They are
not allowed to communicate but have access to the following payoff matrix:

\begin{equation}
\text{Alice}%
\begin{array}{c}
C \\ 
D%
\end{array}%
\overset{\text{{\Large Bob}}}{\overset{%
\begin{array}{cc}
C\text{ \ \ \ \ } & D%
\end{array}%
}{\left[ 
\begin{array}{cc}
\left( 3,3\right) & \left( 0,5\right) \\ 
\left( 5,0\right) & \left( 1,1\right)%
\end{array}%
\right] }}  \tag{$A1$}
\end{equation}%
It is clear from the payoff matrix $A1$ that $D$ is the dominant strategy
for the two players. Therefore, rational reasoning forces the players to
play $D$. Hence ($D,D$) is the Nash equilibrium of the game with payoffs ($%
1,1$). But the players could get higher payoffs if they would have played $C$
instead of $D$. This is the dilemma of the game.

\begin{center}
\textbf{Battle of Sexes}
\end{center}

The payoff matrix for Battle of Sexes game is

\begin{equation}
\text{Alice}%
\begin{array}{c}
O \\ 
T%
\end{array}%
\overset{\text{{\Large Bob}}}{\overset{%
\begin{array}{cc}
O\text{ \ \ \ \ } & T%
\end{array}%
}{\left[ 
\begin{array}{cc}
\left( 2,1\right) & \left( 0,0\right) \\ 
\left( 0,0\right) & \left( 1,2\right)%
\end{array}%
\right] }}  \tag{$A2$}
\end{equation}%
In this game Alice is fond of Opera whereas Bob likes watching TV but they
also want to spend the evening together. The two pure Nash equilibria (NE)
of this game are ($O,O$) and ($T,T$) which corresponds to the situation when
both the players choose Opera and TV, respectively. Here the first NE is
more favorable to Alice while the second NE is favorable to Bob. Since they
are not allowed to communicate, So, they face a dilemma in choosing their
strategies.

\begin{center}
\textbf{The Chicken game}
\end{center}

The payoff matrix for the Chicken game is

\begin{equation}
\text{Alice}%
\begin{array}{c}
C \\ 
D%
\end{array}%
\overset{\text{{\Large Bob}}}{\overset{%
\begin{array}{cc}
C\text{ \ \ \ \ } & D%
\end{array}%
}{\left[ 
\begin{array}{cc}
\left( 3,3\right) & \left( 1,4\right) \\ 
\left( 4,1\right) & \left( 0,0\right)%
\end{array}%
\right] }}  \tag{$A3$}
\end{equation}%
In the game of Chicken, also known as the Hawk-Dove game, two players drove
their cars towards each other. The first one to swerve to avoid collision is
the loser (chicken) and the one who keeps on driving straight is the winner.
There is no dominant strategy in this game. There are two NE ($C,D$) and ($%
D,C$), the former is preferred by Bob and the latter is preferred by Alice.
The dilemma of this game is that the Pareto Optimal strategy ($C,C$) is not
Nash equilibrium.\newpage

\begin{center}
\textbf{Appendix B: Some Definitions}
\end{center}

The definitions of the parameters used in equation (22) are given as

\begin{eqnarray*}
\eta _{1}^{A} &=&\chi _{\mu _{1}}^{(00)}\chi _{\mu _{2}}^{(00)}\cos
^{2}(\gamma /2)\cos ^{2}(\delta /2)+(\sin ^{2}(\gamma /2)+\chi _{\mu
_{1}}^{(11)}\cos ^{2}(\gamma /2))\sin ^{2}(\delta /2) \\
&&+(\chi _{\mu _{1}}^{(00)}\chi _{\mu _{2}}^{(11)}+2\chi _{\mu
_{1}}^{(01)}\chi _{\mu _{2}}^{(a)})\sin {}^{2}(\delta /2)\cos {}^{2}(\gamma
/2) \\
\eta _{2}^{A} &=&\chi _{\mu _{2}}^{(00)}(\sin ^{2}(\gamma /2)+\chi _{\mu
_{1}}^{(11)}\cos ^{2}(\gamma /2))\cos ^{2}(\delta /2)+(\chi _{\mu
_{1}}^{(00)}+2\chi _{\mu _{1}}^{(01)}\chi _{\mu _{2}}^{(a)})\times \\
&&\sin {}^{2}(\delta /2)\cos {}^{2}(\gamma /2)+\chi _{\mu _{2}}^{(11)}(\sin
^{2}(\gamma /2)+\chi _{\mu _{1}}^{(11)}\cos ^{2}(\gamma /2))\sin ^{2}(\delta
/2) \\
\eta _{3}^{A} &=&\chi _{\mu _{1}}^{(01)}\chi _{\mu _{2}}^{(00)}\cos
^{2}(\gamma /2)\cos ^{2}(\delta /2)+(\chi _{\mu _{1}}^{(01)}+\chi _{\mu
_{1}}^{(01)}\chi _{\mu _{2}}^{(11)})\cos ^{2}(\gamma /2)\sin ^{2}(\delta /2)
\\
&&+\chi _{\mu _{2}}^{(a)}(\sin ^{2}(\gamma /2)+\chi _{\mu _{1}}^{(11)}\cos
^{2}(\gamma /2)+\chi _{\mu _{1}}^{(00)}\cos {}^{2}(\gamma /2))\sin
^{2}(\delta /2) \\
\eta _{4}^{A} &=&\chi _{\mu _{1}}^{(10)}\chi _{\mu _{2}}^{(00)}\cos
^{2}(\delta /2)+(\chi _{\mu _{1}}^{(10)}+\chi _{\mu _{1}}^{(10)}\chi _{\mu
_{2}}^{(11)}-2\chi _{\mu _{1}}^{(10)}\chi _{\mu _{2}}^{(a)})\sin ^{2}(\delta
/2)
\end{eqnarray*}%
\begin{eqnarray*}
\chi _{1}^{A} &=&\chi _{\mu _{1}}^{(00)}\chi _{\mu _{2}}^{(00)}\cos
^{2}(\gamma /2)\sin ^{2}(\delta /2)+(\sin ^{2}(\gamma /2)+\chi _{\mu
_{1}}^{(11)}\cos ^{2}(\gamma /2))\cos ^{2}(\delta /2) \\
&&+(\chi _{\mu _{1}}^{(00)}\chi _{\mu _{2}}^{(11)}+2\chi _{\mu
_{1}}^{(01)}\chi _{\mu _{2}}^{(a)})\cos {}^{2}(\delta /2)\cos {}^{2}(\gamma
/2) \\
\chi _{2}^{A} &=&\chi _{\mu _{2}}^{(00)}(\sin ^{2}(\gamma /2)+\chi _{\mu
_{1}}^{(11)}\cos ^{2}(\gamma /2))\sin ^{2}(\delta /2)+(\chi _{\mu
_{1}}^{(00)}+2_{\mu _{1}}^{(01)}\chi _{\mu _{2}}^{(a)})\times \\
&&\cos {}^{2}(\delta /2)\cos {}^{2}(\gamma /2)+\chi _{\mu _{2}}^{(11)}(\sin
^{2}(\gamma /2)+\chi _{\mu _{1}}^{(11)}\cos ^{2}(\gamma /2))\cos ^{2}(\delta
/2) \\
\chi _{3}^{A} &=&(\chi _{\mu _{1}}^{(01)}\chi _{\mu _{2}}^{(00)}+\chi _{\mu
_{1}}^{(01)}\chi _{\mu _{2}}^{(11)})\cos ^{2}(\gamma /2)\sin ^{2}(\delta
/2)+\chi _{\mu _{1}}^{(01)}\cos ^{2}(\gamma /2)\cos ^{2}(\delta /2) \\
&&+\chi _{\mu _{2}}^{(a)}(\sin ^{2}(\gamma /2)+\chi _{\mu _{1}}^{(11)}\cos
^{2}(\gamma /2)+\chi _{\mu _{1}}^{(00)}\cos {}^{2}(\gamma /2))\cos
^{2}(\delta /2) \\
\chi _{4}^{A} &=&\chi _{\mu _{1}}^{(10)}\chi _{\mu _{2}}^{(00)}\sin
^{2}(\delta /2)+(_{\mu _{1}}^{(10)}+\chi _{\mu _{1}}^{(10)}\chi _{\mu
_{2}}^{(11)}-2\chi _{\mu _{1}}^{(10)}\chi _{\mu _{2}}^{(a)})\cos ^{2}(\delta
/2)
\end{eqnarray*}%
\begin{eqnarray*}
\Delta _{1}^{A} &=&(\chi _{\mu _{1}}^{(01)}\chi _{\mu _{2}}^{(b)}+\chi _{\mu
_{1}}^{(00)}\chi _{\mu _{2}}^{(01)})\cos ^{2}(\gamma /2) \\
\Delta _{2}^{A} &=&\chi _{\mu _{1}}^{(01)}\chi _{\mu _{2}}^{(b)}\cos
^{2}(\gamma /2)+\chi _{\mu _{2}}^{(01)}(\sin ^{2}(\gamma /2)+\chi _{\mu
_{1}}^{(11)}\cos ^{2}(\gamma /2)) \\
\Delta _{3}^{A} &=&\chi _{\mu _{2}}^{(b)}(\sin ^{2}(\gamma /2)+\chi _{\mu
_{1}}^{(11)}\cos ^{2}(\gamma /2))\cos ^{2}(\delta /2) \\
&&+(\chi _{\mu _{1}}^{(01)}\chi _{\mu _{2}}^{(01)}+\chi _{\mu
_{1}}^{(00)}\chi _{\mu _{2}}^{(b)}\sin ^{2}(\delta /2))\cos ^{2}(\gamma /2)
\\
\Delta _{4}^{A} &=&\chi _{\mu _{2}}^{(b)}(\sin ^{2}(\gamma /2)+\chi _{\mu
_{1}}^{(11)}\cos ^{2}(\gamma /2))\sin ^{2}(\delta /2) \\
&&+(\chi _{\mu _{1}}^{(01)}\chi _{\mu _{2}}^{(01)}+\chi _{\mu
_{1}}^{(00)}\chi _{\mu _{2}}^{(b)}\cos ^{2}(\delta /2))\cos ^{2}(\gamma /2)
\\
\Delta _{5}^{A} &=&(\sin ^{2}(\gamma /2)+\chi _{\mu _{1}}^{(11)}\cos
^{2}(\gamma /2))+\chi _{\mu _{1}}^{(00)}-2\chi _{\mu _{1}}^{(01)}\cos
^{2}(\gamma /2),\quad \\
\Delta _{6}^{A} &=&\chi _{\mu _{1}}^{(10)}\chi _{\mu _{2}}^{(01)}-_{\mu
_{1}}^{(10)}\chi _{\mu _{2}}^{(b)},\text{ }\xi =\frac{1}{2}\sin (\delta
)\sin (\gamma ), \\
s_{i} &=&\sin {}^{2}(\frac{\theta _{i}}{2}),\text{\quad }c_{i}=\cos ^{2}(%
\frac{\theta _{i}}{2})
\end{eqnarray*}%
\begin{eqnarray*}
\chi _{\mu _{1}}^{(00)} &=&(1-p_{1})^{2}+\mu _{1}(1-p_{1})p_{1},\quad \\
\chi _{\mu 2}^{(00)} &=&(1-p_{2})^{2}+\mu _{2}(1-p_{2})p_{2} \\
\chi _{\mu _{1}}^{(11)} &=&p_{1}{}^{2}+\mu _{1}(1-p_{1})p_{1},\quad \\
\chi _{\mu _{2}}^{(11)} &=&p_{2}{}^{2}+\mu _{2}(1-p_{2})p_{2} \\
\chi _{\mu _{1}}^{(10)} &=&(1-\mu _{1})(1-p_{1})+\mu _{1}(1-p_{1})^{\frac{1}{%
2}},\quad \\
\chi _{\mu 2}^{(10)} &=&(1-\mu _{2})(1-p_{2})+\mu _{2}(1-p_{2})^{\frac{1}{2}}
\\
\chi _{\mu _{1}}^{(01)} &=&(1-\mu _{1})(1-p_{1})p_{1},\quad \chi _{\mu
2}^{(01)}=(1-\mu _{2})(1-p_{2})p_{2} \\
\chi _{\mu _{2}}^{(a)} &=&(1-\mu _{2})p_{2},\quad \chi _{\mu
2}^{(b)}=(1-p_{2})+\mu _{2}p_{2}
\end{eqnarray*}%
The definitions of the parameters used in equation (23) are given as%
\begin{eqnarray*}
\Delta _{\mu 1}^{1} &=&-\frac{1}{9}\left( -3+2p_{1}\right) \left(
-2p_{1}+2\mu _{1}p_{1}+3\right) \\
\Delta _{\mu 1}^{2} &=&-\frac{2}{9}p_{1}\left( -2p_{1}+2\mu _{1}p_{1}-3\mu
_{1}\right) \\
\Delta _{\mu 1}^{3} &=&-\frac{1}{9}\left( -9+24p_{1}-18\mu
_{1}p_{1}-16p_{1}^{2}+16\mu _{1}p_{1}^{2}\right) -\frac{2}{3}\mu _{1}p_{1} \\
\Delta _{\mu 1}^{4} &=&\frac{2}{9}p_{1}\left( -3+2p_{1}\right) \left( \mu
_{1}-1\right) \\
\Delta _{\mu 2}^{1} &=&-\frac{1}{9}\left( -3+2p_{2}\right) \left(
-2p_{2}+2\mu _{2}p_{2}+3\right) \\
\Delta _{\mu 2}^{2} &=&\frac{2}{9}p_{2}\left( -3+2p_{2}\right) \left( \mu
_{2}-1\right) \\
\Delta _{\mu 2}^{3} &=&-\frac{2}{9}p_{2}\left( -2p_{2}+2\mu _{2}p_{2}-3\mu
_{2}\right) \\
\Delta _{\mu 2}^{4} &=&-\frac{1}{9}\left( -9+24p_{2}-18\mu
_{2}p_{2}-16p_{2}^{2}+16\mu _{2}p_{2}^{2}\right) \\
\Delta _{\mu }^{11} &=&\Delta _{\mu 1}^{1}\cos ^{2}(\gamma /2)+\Delta _{\mu
1}^{2}\sin ^{2}(\gamma /2) \\
\Delta _{\mu }^{21} &=&\Delta _{\mu 1}^{2}\cos ^{2}(\gamma /2)+\Delta _{\mu
1}^{1}\sin ^{2}(\gamma /2)
\end{eqnarray*}%
\begin{eqnarray*}
\eta ^{D} &=&(\Delta _{\mu 2}^{1}\Delta _{\mu }^{11}+\Delta _{\mu
2}^{3}\Delta _{\mu }^{21})\cos ^{2}(\delta /2)+(\Delta _{\mu 2}^{1}\Delta
_{\mu }^{21}+\Delta _{\mu 2}^{3}\Delta _{\mu }^{11})\sin ^{2}(\delta
/2)+2\Delta _{\mu 2}^{2}\Delta _{\mu 1}^{4} \\
\chi ^{D} &=&(\Delta _{\mu 2}^{1}\Delta _{\mu }^{21}+\Delta _{\mu
2}^{3}\Delta _{\mu }^{11})\cos ^{2}(\delta /2)+(\Delta _{\mu 2}^{1}\Delta
_{\mu }^{11}+\Delta _{\mu 2}^{3}\Delta _{\mu }^{21})\sin ^{2}(\delta
/2)+2\Delta _{\mu 2}^{2}\Delta _{\mu 1}^{4} \\
\Delta ^{D} &=&\Delta _{\mu 2}^{2}\Delta _{\mu }^{11}+\Delta _{\mu
2}^{2}\Delta _{\mu }^{21}+\Delta _{\mu 2}^{1}\Delta _{\mu 1}^{4}+\Delta
_{\mu 2}^{3}\Delta _{\mu 1}^{4} \\
\eta _{1DP} &=&-(\Delta _{\mu 1}^{2}\cos ^{2}(\gamma /2)+\Delta _{\mu
1}^{1}\sin ^{2}(\gamma /2))-(\Delta _{\mu 1}^{2}\sin ^{2}(\gamma /2)+\Delta
_{\mu 1}^{1}\cos ^{2}(\gamma /2))+2\Delta _{\mu 1}^{4}
\end{eqnarray*}%
The definitions of the parameters used in equation (24) are given as%
\begin{eqnarray*}
\eta ^{P} &=&\cos ^{2}(\gamma /2)\cos ^{2}(\delta /2)+\sin ^{2}(\gamma
/2)\sin ^{2}(\delta /2) \\
\chi ^{P} &=&\sin ^{2}(\gamma /2)\cos ^{2}(\delta /2)+\cos ^{2}(\gamma
/2)\sin ^{2}(\delta /2) \\
\mu _{p}^{(i)} &=&(1-\mu _{i})(1-p_{i})^{2}+\mu _{i}
\end{eqnarray*}%
The definitions of the parameters used in equation (25) are given as%
\begin{eqnarray*}
\eta _{1}^{PA} &=&\chi _{\mu 2}^{(00)}\cos ^{2}(\gamma /2)\cos ^{2}(\delta
/2)+(\sin ^{2}(\gamma /2)+\chi _{\mu 2}^{(11)}\cos ^{2}(\gamma /2))\sin
^{2}(\delta /2) \\
\eta _{2}^{PA} &=&(\cos ^{2}(\gamma /2)+\chi _{\mu 2}^{(11)}\sin ^{2}(\gamma
/2))\sin ^{2}(\delta /2)+\chi _{\mu 2}^{(00)}\sin ^{2}(\gamma /2)\cos
^{2}(\delta /2) \\
\eta _{3}^{PA} &=&\chi _{\mu 2}^{(a)}\sin ^{2}(\gamma /2)\sin ^{2}(\delta
/2)+\chi _{\mu 2}^{(a)}\cos ^{2}(\gamma /2)\sin ^{2}(\delta /2) \\
\chi _{1}^{PA} &=&(\cos ^{2}(\gamma /2)+\chi _{\mu 2}^{(11)}\sin ^{2}(\gamma
/2))\cos ^{2}(\delta /2)+\chi _{\mu 2}^{(00)}\cos ^{2}(\gamma /2)\sin
^{2}(\delta /2) \\
\chi _{2}^{PA} &=&\chi _{\mu 2}^{(00)}\sin ^{2}(\gamma /2)\sin ^{2}(\delta
/2)+(\cos ^{2}(\gamma /2)+\chi _{\mu 2}^{(11)}\sin ^{2}(\gamma /2))\cos
^{2}(\delta /2) \\
\chi _{3}^{PA} &=&\chi _{\mu 2}^{(a)}\sin ^{2}(\gamma /2)\cos ^{2}(\delta
/2)+\chi _{\mu 2}^{(a)}\cos ^{2}(\gamma /2)\cos ^{2}(\delta /2) \\
\Delta _{1}^{PA} &=&\chi _{\mu 2}^{(01)}\cos ^{2}(\gamma /2)\cos ^{2}(\delta
/2)+\chi _{\mu 2}^{(01)}\cos ^{2}(\gamma /2)\sin ^{2}(\delta /2) \\
\Delta _{2}^{PA} &=&\chi _{\mu 2}^{(01)}\sin ^{2}(\gamma /2)\cos ^{2}(\delta
/2)+\chi _{\mu 2}^{(01)}\sin ^{2}(\gamma /2)\sin ^{2}(\delta /2) \\
\Delta _{3}^{PA} &=&\chi _{\mu 2}^{(b)}\sin ^{2}(\gamma /2)\cos ^{2}(\delta
/2)+\chi _{\mu 2}^{(b)}\cos ^{2}(\gamma /2)\sin ^{2}(\delta /2) \\
\Delta _{4}^{PA} &=&\chi _{\mu 2}^{(b)}\cos ^{2}(\gamma /2)\cos ^{2}(\delta
/2)+\chi _{\mu 2}^{(b)}\sin ^{2}(\gamma /2)\sin ^{2}(\delta /2)
\end{eqnarray*}%
The definitions of the parameters used in equation (26) are given as%
\begin{eqnarray*}
\eta ^{AP} &=&\chi _{\mu 1}^{(00)}\cos ^{2}(\gamma /2)\cos ^{2}(\delta
/2)+(\sin ^{2}(\gamma /2)+\chi _{\mu 1}^{(11)}\cos ^{2}(\gamma /2))\sin
^{2}(\delta /2) \\
\chi ^{AP} &=&(\sin ^{2}(\gamma /2)+\chi _{\mu 1}^{(11)}\cos ^{2}(\gamma
/2))\cos ^{2}(\delta /2)+\chi _{\mu 1}^{(00)}\cos ^{2}(\gamma /2)\sin
^{2}(\delta /2) \\
\Delta ^{AP} &=&\chi _{\mu 1}^{(01)}\cos ^{2}(\gamma /2)\cos ^{2}(\delta
/2)+\chi _{\mu 1}^{(01)}\cos ^{2}(\gamma /2)\sin ^{2}(\delta /2)
\end{eqnarray*}%
The definitions of the parameters used in equation (27) are given as%
\begin{eqnarray*}
\eta ^{AD} &=&(\chi _{\mu 1}^{(00)}\Delta _{\mu 2}^{1}\cos ^{2}(\gamma
/2)+\Delta _{\mu 2}^{3}(\sin ^{2}(\gamma /2)+\chi _{\mu 1}^{(11)}\cos
^{2}(\gamma /2)))\cos ^{2}(\delta /2)+(\Delta _{\mu 2}^{1}(\sin ^{2}(\gamma
/2) \\
&&+\chi _{\mu 1}^{(11)}\cos ^{2}(\gamma /2))+\chi _{\mu 1}^{(00)}\Delta
_{\mu 2}^{3}\cos ^{2}(\gamma /2))\sin ^{2}(\delta /2)+2\chi _{\mu
1}^{(01)}\Delta _{\mu 2}^{2}\cos ^{2}(\gamma /2) \\
\chi ^{AD} &=&(\chi _{\mu 1}^{(00)}\Delta _{\mu 2}^{1}\cos ^{2}(\gamma
/2)+\Delta _{\mu 2}^{3}(\sin ^{2}(\gamma /2)+\chi _{\mu 1}^{(11)}\cos
^{2}(\gamma /2)))\sin ^{2}(\delta /2)+(\Delta _{\mu 2}^{1}(\sin ^{2}(\gamma
/2) \\
&&+\chi _{\mu 1}^{(11)}\cos ^{2}(\gamma /2))+\chi _{\mu 1}^{(00)}\Delta
_{\mu 2}^{3}\cos ^{2}(\gamma /2))\cos ^{2}(\delta /2)+2\chi _{\mu
1}^{(01)}\Delta _{\mu 2}^{2}\cos ^{2}(\gamma /2) \\
\Delta ^{AD} &=&\chi _{\mu 1}^{(01)}(\Delta _{\mu 2}^{1}+\Delta _{\mu
2}^{3})\cos ^{2}(\gamma /2)+\Delta _{\mu 2}^{2}(\sin ^{2}(\gamma /2)+\chi
_{\mu 1}^{(11)}\cos ^{2}(\gamma /2))+\chi _{\mu 1}^{(00)}\Delta _{\mu
2}^{2}\cos ^{2}(\gamma /2) \\
\eta _{1AD} &=&\chi _{\mu 1}^{(11)}\cos ^{2}(\gamma /2))+\chi _{\mu
1}^{(00)}\cos ^{2}(\gamma /2)-2\chi _{\mu 1}^{(01)}\cos ^{2}(\gamma /2)+\sin
^{2}(\gamma /2)
\end{eqnarray*}%
The definitions of the parameters used in equation (28) are given as%
\begin{eqnarray*}
\eta _{1}^{DA} &=&\Delta _{\mu }^{11}\chi _{\mu 2}^{(00)}\cos ^{2}(\delta
/2)+\Delta _{\mu }^{11}\chi _{\mu 2}^{(11)}\sin ^{2}(\delta /2)+\Delta _{\mu
}^{12}\sin ^{2}(\delta /2)+2\Delta _{\mu 1}^{4}\chi _{\mu 2}^{(a)}\sin
^{2}(\delta /2) \\
\eta _{2}^{DA} &=&\Delta _{\mu }^{21}\chi _{\mu 2}^{(00)}\cos ^{2}(\delta
/2)+\Delta _{\mu }^{21}\chi _{\mu 2}^{(11)}\sin ^{2}(\delta /2)+\Delta _{\mu
}^{11}\sin ^{2}(\delta /2)+2\Delta _{\mu 1}^{4}\chi _{\mu 2}^{(a)}\sin
^{2}(\delta /2) \\
\eta _{3}^{DA} &=&(\Delta _{\mu }^{11}\chi _{\mu 2}^{(a)}+\Delta _{\mu
}^{21}\chi _{\mu 2}^{(a)})\sin ^{2}(\delta /2)+\Delta _{\mu 1}^{4}\chi _{\mu
2}^{(00)}\cos ^{2}(\delta /2)+(\chi _{\mu 2}^{(11)}+1)\Delta _{\mu
1}^{4}\sin ^{2}(\delta /2) \\
\chi _{1}^{DA} &=&\Delta _{\mu }^{11}\chi _{\mu 2}^{(00)}\sin ^{2}(\delta
/2)+\Delta _{\mu }^{11}\chi _{\mu 2}^{(11)}\cos ^{2}(\delta /2)+\Delta _{\mu
}^{12}\cos ^{2}(\delta /2)+2\Delta _{\mu 1}^{4}\chi _{\mu 2}^{(a)}\cos
^{2}(\delta /2) \\
\chi _{2}^{DA} &=&\Delta _{\mu }^{21}\chi _{\mu 2}^{(00)}\sin ^{2}(\delta
/2)+\Delta _{\mu }^{21}\chi _{\mu 2}^{(11)}\cos ^{2}(\delta /2)+\Delta _{\mu
}^{11}\cos ^{2}(\delta /2)+2\Delta _{\mu 1}^{4}\chi _{\mu 2}^{(a)}\cos
^{2}(\delta /2) \\
\chi _{3}^{DA} &=&(\Delta _{\mu }^{11}\chi _{\mu 2}^{(a)}+\Delta _{\mu
}^{21}\chi _{\mu 2}^{(a)})\cos ^{2}(\delta /2)+\Delta _{\mu 1}^{4}\chi _{\mu
2}^{(00)}\sin ^{2}(\delta /2)+(\chi _{\mu 2}^{(11)}+1)\Delta _{\mu
1}^{4}\cos ^{2}(\delta /2) \\
\Delta _{1}^{DA} &=&\Delta _{\mu }^{11}\chi _{\mu 2}^{(01)}+\Delta _{\mu
1}^{4}\chi _{\mu 2}^{(b)} \\
\Delta _{2}^{DA} &=&\Delta _{\mu }^{11}\chi _{\mu 2}^{(b)}\sin ^{2}(\delta
/2)+\Delta _{\mu }^{21}\chi _{\mu 2}^{(b)}\cos ^{2}(\delta /2)+\Delta _{\mu
1}^{4}\chi _{\mu 2}^{(01)} \\
\Delta _{3}^{DA} &=&\Delta _{\mu }^{11}\chi _{\mu 2}^{(b)}\cos ^{2}(\delta
/2)+\Delta _{\mu }^{21}\chi _{\mu 2}^{(b)}\sin ^{2}(\delta /2)+\Delta _{\mu
1}^{4}\chi _{\mu 2}^{(01)}
\end{eqnarray*}%
The definitions of the parameters used in equation (29) are given as%
\begin{eqnarray*}
\eta ^{DP} &=&\Delta _{\mu }^{11}\cos ^{2}(\delta /2)+\Delta _{\mu
}^{12}\sin ^{2}(\delta /2) \\
\chi ^{DP} &=&\Delta _{\mu }^{11}\sin ^{2}(\delta /2)+\Delta _{\mu
}^{12}\cos ^{2}(\delta /2)
\end{eqnarray*}%
The definitions of the parameters used in equation (30) are given as%
\begin{eqnarray*}
\eta ^{PD} &=&(\Delta _{\mu 2}^{1}\cos ^{2}(\gamma /2)+\Delta _{\mu
2}^{3}\sin ^{2}(\gamma /2))\cos ^{2}(\delta /2) \\
&&+(\Delta _{\mu 2}^{1}\sin ^{2}(\gamma /2)+\Delta _{\mu 2}^{3}\cos
^{2}(\gamma /2))\sin ^{2}(\delta /2) \\
\chi ^{PD} &=&(\Delta _{\mu 2}^{1}\sin ^{2}(\gamma /2)+\Delta _{\mu
2}^{3}\cos ^{2}(\gamma /2))\cos ^{2}(\delta /2) \\
&&+(\Delta _{\mu 2}^{1}\cos ^{2}(\gamma /2)+\Delta _{\mu 2}^{3}\sin
^{2}(\gamma /2))\sin ^{2}(\delta /2)
\end{eqnarray*}

\begin{figure}[tbp]
\begin{center}
\vspace{-2cm} \includegraphics[scale=0.6]{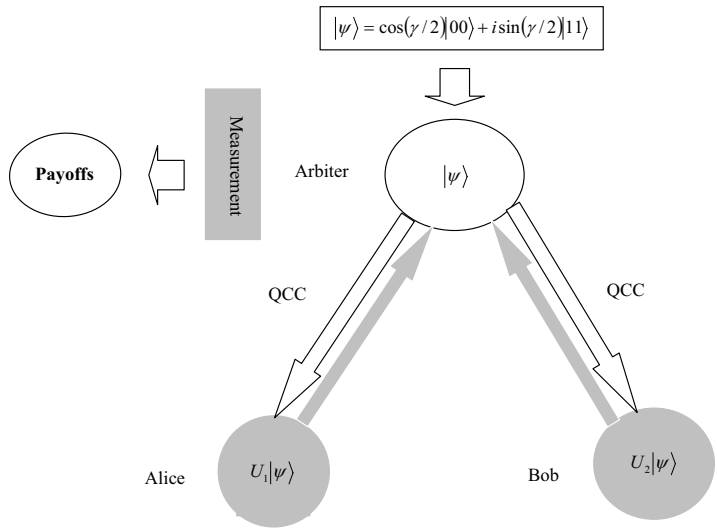} \\[0pt]
\end{center}
\caption{Schematic diagram of the model.}
\end{figure}

\begin{figure}[tbp]
\begin{center}
\vspace{-2cm} \includegraphics[scale=0.6]{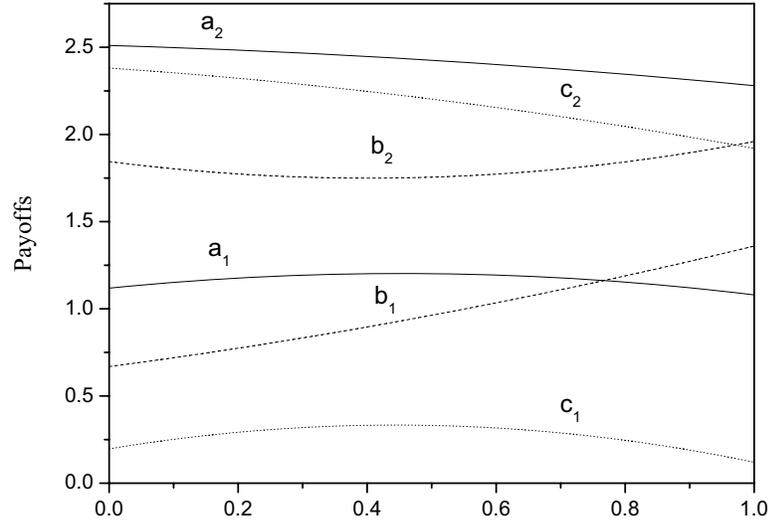} \\[0pt]
\end{center}
\caption{Players (Alice/Bob) payoffs as a function of the memory parameter, $%
\protect\mu $ is plotted for the quantum games Prisoner's Dilemma (solid
lines, $a_{i}$), Battle of the Sexes (dashed lines, $b_{i}$) and Chicken
(dotted lines, $c_{i}$) for amplitude-damping channel. Indices 1 \& 2
correspond to $p=0.8$ \& $p=0.2$ respectively with $\protect\delta =\protect%
\gamma =0,$ $\protect\theta _{1}=0,$ $\protect\theta _{2}=\protect\pi /2$
and $\protect\alpha _{2}=\protect\pi /2,$ $\protect\beta _{2}=0$ as Bob's
optimal strategy.}
\end{figure}

\begin{figure}[tbp]
\begin{center}
\vspace{-2cm} \includegraphics[scale=0.6]{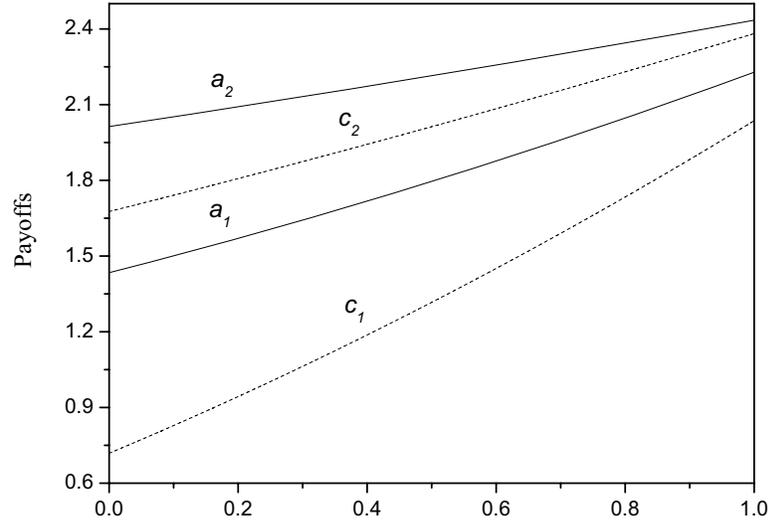} \\[0pt]
\end{center}
\caption{Bob's payoff as a function of memory parameter $\protect\mu $ is
plotted for Prisoner's Dilemma (solid lines, $a_{i}$) and Chicken (dashed
lines, $c_{i}$) for amplitude-damping channel. Indices 1 \& 2 correspond to $%
p=0.8$ \& $p=0.2$ respectively with $\protect\delta =0,$ $\protect\gamma =%
\protect\pi /2,$ $\protect\theta _{1}=\protect\theta _{2}=\protect\pi /2$
and $\protect\alpha _{2}=\protect\pi /2,$ $\protect\beta _{2}=0$ as his
optimal strategy.}
\end{figure}

\begin{figure}[tbp]
\begin{center}
\vspace{-2cm} \includegraphics[scale=0.6]{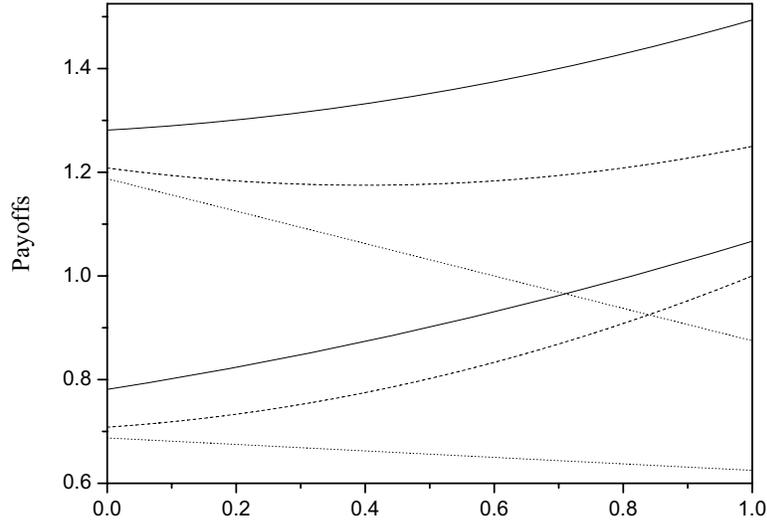} \\[0pt]
\end{center}
\caption{Payoffs for Alice (classical player) and Bob (quantum player) are
plotted as a function of memory parameter $\protect\mu $ for
amplitude-damping (solid lines), depolarizing followed by an
amplitude-damping (dashed lines) and phase-damping followed by an
amplitude-damping (dotted lines) channels for Battle of the Sexes game with $%
\protect\delta =0,$ $\protect\gamma =\protect\pi /2,$ $p=0.5,$ $\protect%
\theta _{1}=0,$ $\protect\theta _{2}=\protect\pi /2$ and $\protect\alpha %
_{2}=\protect\pi /2,$ $\protect\beta _{2}=0$ as Bob's optimal strategy. The
lower curves for all the three cases correspond to Alice's payoff.}
\end{figure}

\begin{figure}[tbp]
\begin{center}
\vspace{-2cm} \includegraphics[scale=0.6]{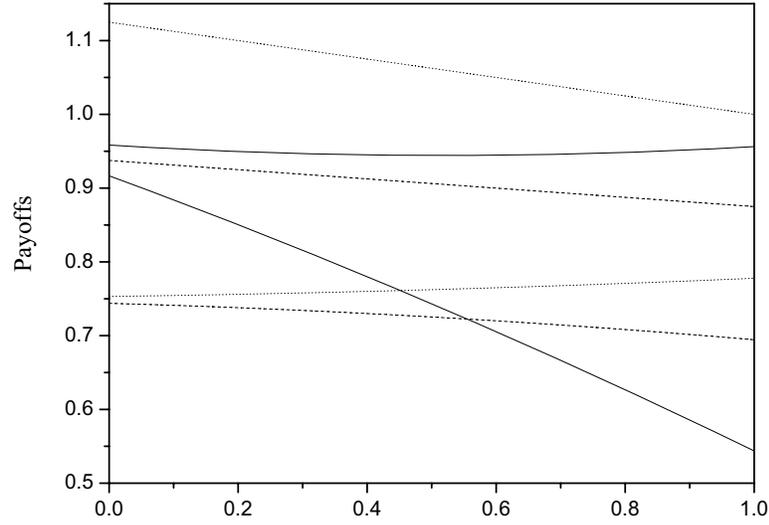} \\[0pt]
\end{center}
\caption{Payoffs for Alice and Bob are plotted as a function of memory
parameter $\protect\mu $ for amplitude-damping (solid lines), depolarizing
followed by an amplitude-damping (dashed lines) and phase-damping followed
by an amplitude-damping (dotted lines) channels for Battle of the Sexes game
with $\protect\gamma =0,$ $\protect\delta =\protect\pi /2,$ $p=0.5,$ $%
\protect\theta _{1}=0,$ $\protect\theta _{2}=\protect\pi /2$ and $\protect%
\alpha _{2}=0,$ $\protect\beta _{2}=\protect\pi /2$ as Bob's optimal
strategy. The lower curves for all the three cases correspond to Alice's
payoff.}
\end{figure}

\begin{figure}[tbp]
\begin{center}
\vspace{-2cm} \includegraphics[scale=0.6]{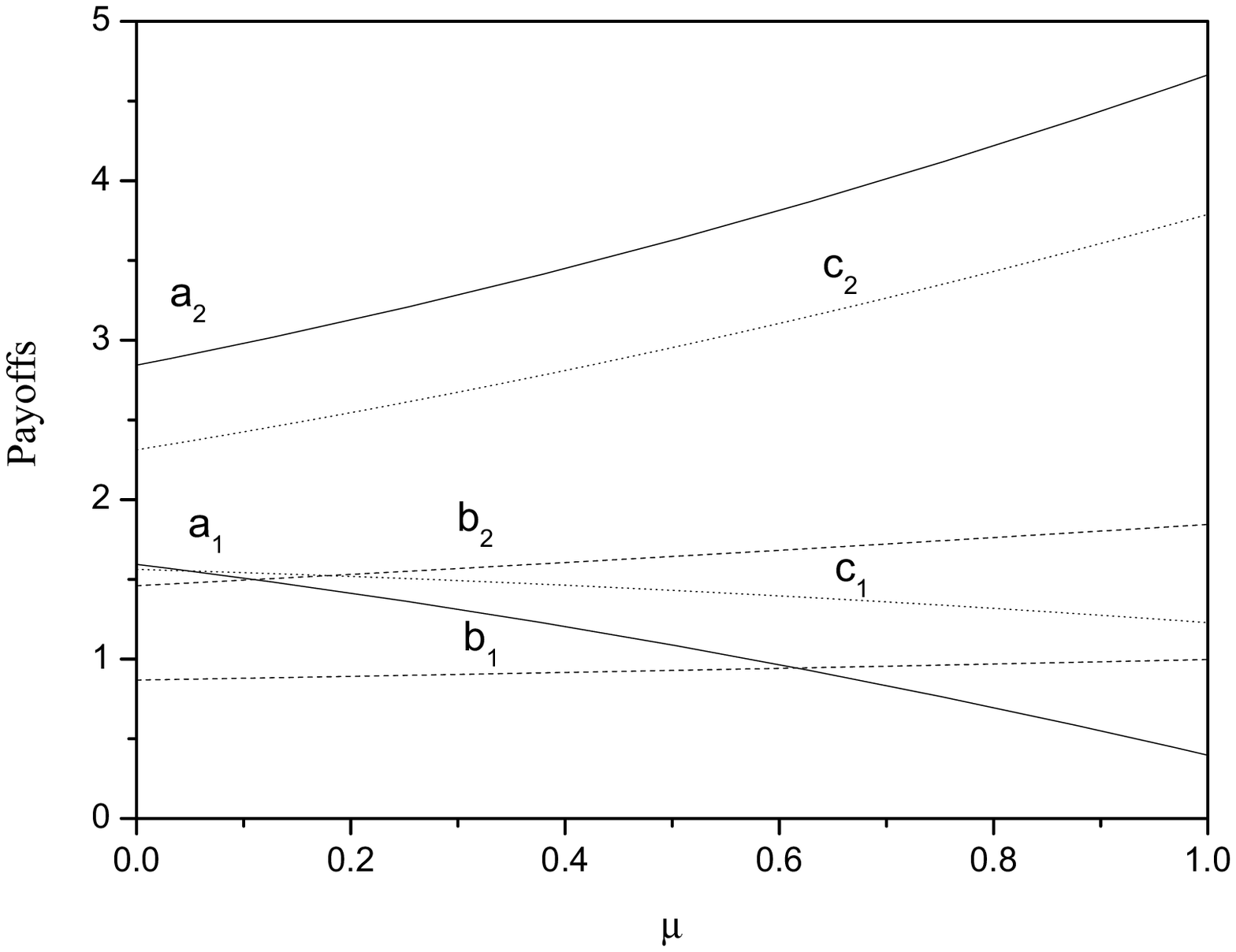} \\[0pt]
\end{center}
\caption{ Alice's (1) and Bob's (2) payoffs are plotted as a function of
memory parameter $\protect\mu $ for the quantum games Prisoner's Dilemma
(solid lines, $a_{i}$), Battle of the Sexes (dashed lines, $b_{i}$) and
Chicken (dotted lines, $c_{i}$) for amplitude-damping channel with $\protect%
\delta =\protect\gamma =\protect\pi /2,$ $p=0.5,$ $\protect\theta _{1}=0,$ $%
\protect\theta _{2}=\protect\pi /2$ and $\protect\alpha _{2}=\protect\pi /2,$
$\protect\beta _{2}=0$ as Bob's optimal strategy.}
\end{figure}

\begin{figure}[tbp]
\begin{center}
\vspace{-2cm} \includegraphics[scale=0.6]{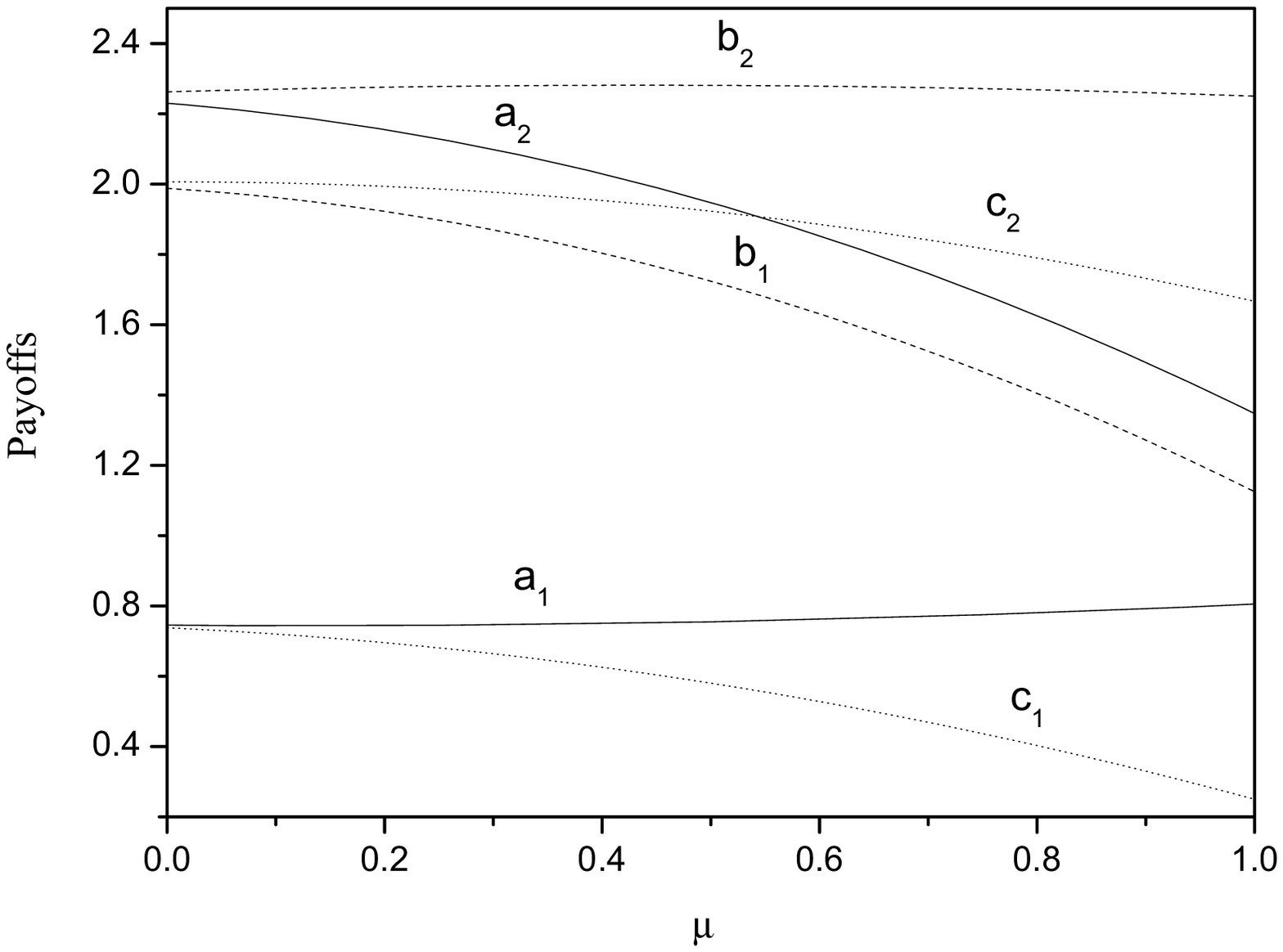} \\[0pt]
\end{center}
\caption{Alice's (1) and Bob's (2) payoffs are plotted as a function of the
memory parameter $\protect\mu $ for the quantum games Prisoner's Dilemma
(solid lines, $a_{i}$), Battle of the Sexes (dashed lines, $b_{i}$) and
Chicken (dotted lines, $c_{i}$) for depolarizing channel with $\protect%
\delta =\protect\gamma =\protect\pi /2,$ $p=0.5,$ $\protect\theta _{1}=0,$ $%
\protect\theta _{2}=\protect\pi /2$ and $\protect\alpha _{2}=\protect\pi /2,$
$\protect\beta _{2}=0$ as Bob's optimal strategy.}
\end{figure}

\end{document}